\documentclass[10pt]{article}
\usepackage[T1]{fontenc}
\usepackage[latin9]{inputenc}
\usepackage{xcolor}
\usepackage{pdfcolmk}
\usepackage{amsmath}
\usepackage{amssymb}
\usepackage{esint}
\PassOptionsToPackage{normalem}{ulem}
\usepackage{ulem}
\usepackage[unicode=true]
 {hyperref}

\makeatletter

\providecolor{lyxadded}{rgb}{0,0,1}
\providecolor{lyxdeleted}{rgb}{1,0,0}
\usepackage{enumitem}		% customizable list environments
\usepackage{M12}

\hypersetup{citecolor={darkblue},urlcolor={darkblue} }

\titlelabel{\makebox[7mm][l]{\thetitle}}

\renewcommand{\vec}[1]{\oldvec{#1}}
\renewcommand{\bar}[1]{\mkern 1mu\overline{\mkern-1mu#1\mkern-1.5mu}\mkern 1.5mu}

\renewcommand{\hat}[1]{\mkern 2mu\widehat{\mkern-1mu#1\mkern-1mu}\mkern 2mu}
 
\renewcommand{\tilde}[1]{\mkern 2mu\widetilde{\mkern-1mu#1\mkern-1mu}\mkern 2mu}

\definecolor{midblue}{rgb}{0.55, 0.55, 0.9}

\usepackage{slashed}
\renewcommand{\not}[1]{\slashed{#1}}

\newcommand{\ket}[1]{\vert #1 \rangle}

\AtBeginDocument{
  
}

\makeatother

\begin{document}
%% some maths definitions

%\newcommand{\cJ}{\mathcal{J}}
%\newcommand{\cS}{\mathcal{S}}
\newcommand{\cJ}{\scalebox{0.95}{$\mathcal{J}$}}%% causes problems with \frac{1}{\cJ} but otherwise better
\newcommand{\cS}{\scalebox{0.95}{$\mathcal{S}$}}

%% marking quantities defined like the ones in HL
%\newcommand{\ghl}{g_{\text{\tiny{HL}}}}
%\newcommand{\rhohl}{\rho_{\text{\tiny{HL}}}}
%\newcommand{\ehl}{\epsilon_\text{\tiny HL}}
%\newcommand{\xhl}{x_{\text{\tiny{HL}}}}
%\newcommand{\zhl}{z_{\text{\tiny{HL}}}}

\newcommand{\ghl}{\tilde{g}}
\newcommand{\ehl}{\tilde{\epsilon}}
\newcommand{\xhl}{\tilde{x}}
\newcommand{\zhl}{\tilde{z}}
\newcommand{\Qhl}{\tilde{Q}}

\newcommand{\Gsuper}[1]{{G^{(#1)}}}

\newcommand{\qstar}{q_{\star}} 

\newcommand{\pbmn}{\tilde{p}}

\newcommand{\twobw}{\bullet\circ}  %% just like BOSS
\newcommand{\twowb}{\circ\bullet}
\newcommand{\twobb}{\bullet\bullet}
\newcommand{\twoww}{\circ\circ}

\newcommand{\bosst}{\text{\scriptsize{BOSST}}}
\newcommand{\blmt}{\text{\scriptsize{BLMT}}}

\newcommand{\oneover}[1]{\scalebox{0.85}{\raisebox{1mm}{1}\hspace{-0.5mm}\big/$#1$}}

\newcommand{\fudge}{\smash{\raisebox{-2mm}{\includegraphics[width=6mm]{fudge.jpg}}}}

\settitlesize{16.5pt} 
\setauthorsize{12pt}

\title{Massless L\"{u}scher Terms and the Limitations of the $AdS_{3}$
Asymptotic Bethe Ansatz}

\author{Michael C. Abbott\alabel{1} and\hspace{1mm} In\^{e}s Aniceto\alabel{2}
\address[1]{Department of Mathematics, University of Cape Town,\\
Rondebosch 7701, Cape Town, South Africa.} \address[2]{Institute
of Physics, Jagiellonian University, \\
Ul. \L{}ojasiewicza 11, 30-348 Krak\'{o}w, Poland.} \address{michael.abbott@uct.ac.za,
ines@th.if.uj.edu.pl }}

\date{December 2015\\
\href{http://arXiv.org/abs/1512.08761}{arXiv:1512.08761}}
\maketitle
\begin{abstract}
In $AdS_{5}$/CFT$_{4}$ integrability the Bethe ansatz gives the
spectrum of long strings, accurate up to exponentially small corrections.
This is no longer true in $AdS_{3}$, as we demonstrate here by studying
L\"{u}scher F-terms with a massless particle running in the loop.
We apply this to the classic test of Hern\'{a}ndez \& L\'{o}pez,
in which the $su(2)$ sector Bethe equations (including one-loop dressing
phase) should match the semiclassical string theory result for a circular
spinning string. These calculations did not agree in $AdS_{3}\times S^{3}\times T^{4}$,
and we show that the sum of all massless L\"{u}scher F-terms can
reproduce the difference. 
\end{abstract}
\vspace{-8mm}

\tableofcontents{}

\section{Introduction}

There has been much recent work on extending what we have learned
about integrability of strings in $AdS_{5}\times S^{5}$ \cite{Minahan:2002ve,Beisert:2010jr}
to the less than maximally supersymmetric background $AdS_{3}\times S^{3}\times T^{4}$
\cite{Babichenko:2009dk,Sfondrini:2014via}, which arises from a D1-D5
brane intersection. Some results for the massive sector can be adapted
quite simply from the $AdS_{5}$ case, or even generalised to every
$AdS_{n}\times S^{n}$. What is completely new is the presence of
a massless sector , corresponding to the $T^{4}$ directions and their
superpartners. 

In all results to date it has been possible to ignore the massless
excitations when studying the massive sector. This is true for the
calculation of the exact S-matrix from centrally extended symmetries
\cite{Borsato:2013qpa} and the corresponding Bethe equations, for
coset methods \cite{Zarembo:2010yz} and semiclassical energy corrections
to spinning strings \cite{Beccaria:2012kb,Beccaria:2012pm}, for near-BMN
diagrammatic calculations of two- and four-point functions \cite{Sundin:2014sfa,Roiban:2014cia,Sundin:2015uva}\footnote{Earlier papers had to include massless modes on internal legs of diagrams
\cite{Sundin:2013ypa}, but better ways to handle divergent integrals
eventually removed this \cite{Sundin:2014sfa,Roiban:2014cia}. } and one-loop S-matrix via unitarity methods \cite{Bianchi:2013nra,Bianchi:2014rfa},
and for the various calculations of the massive dressing phase \cite{David:2010yg,Borsato:2013hoa,Abbott:2013mpa}. 

But not everything works perfectly in this decoupled picture of the
massive sector, and in particular, the comparison of the one-loop
energy correction to circular spinning strings in $S^{3}$ to the
expansion of the $su(2)$ sector Bethe equations fails. This is precisely
the comparison from which Hern\'{a}ndez and L\'{o}pez constructed
the complete one-loop dressing phase in $AdS_{5}\times S^{5}$ \cite{Beisert:2005cw,Hernandez:2006tk},
and a similar construction in $AdS_{3}\times S^{3}\times T^{4}$ was
done by \cite{Beccaria:2012kb}. However the dressing phase constructed
this way does not solve the crossing equations \cite{Borsato:2013hoa},
nor does it agree with a construction from semiclassical magnon scattering
\cite{Abbott:2013mpa}, nor give the amplitude seen in near-BMN scattering
\cite{Sundin:2014sfa}. Conversely, using the correct dressing phase
(as agreed on by \cite{Borsato:2013hoa}, \cite{Abbott:2013mpa} and
\cite{Sundin:2014sfa}) breaks the circular spinning string comparison.
The resulting mismatch is of the same order as the one-loop energy.

Our paper offers a solution to this problem. L\"{u}scher F-terms
involving massive modes running in the loop give exponentially small
finite-size corrections in $AdS_{5}\times S^{5}$ \cite{Gromov:2008ie,Janik:2010kd}.
But massless modes are in some sense `infinite-range', and thus give
rise to L\"{u}scher F-terms which are not exponentially suppressed.
In fact they are of the same order as the prediction from the asymptotic
Bethe equations, and thus cannot be ignored, even when $L$ is large.
This is the first time it is necessary to include the massless particles
in order to understand the massive sector.

In general one can consider L\"{u}scher corrections wrapping the
space any number of times \cite{Heller:2008at}, and with a massless
virtual particle these all contribute at the same order. For multiparticle
physical states (such as the circular string, with order $\sqrt{\lambda}$
excitations) however only the singly wrapped L\"{u}scher terms are
well-understood \cite{Bajnok:2008bm}. Combining results of \cite{Heller:2008at}
and \cite{Bajnok:2008bm}, we write down a simple formula for a multiparticle
correction wrapping $n$ times. While we believe this omits some multiple
wrapping effects which should contribute at the same order, when applied
to the circular string (and summed over all $n$) it gives the correct
functional form of the mismatch, up to a factor of 2.

\section{Conflicting Results about $su(2)$ Circular Strings\label{sec:The-Problem}}

The $su(2)$ sector at strong coupling concerns strings in $\mathbb{R}\times S^{3}$.
For classical string theory this is a consistent truncation of both
$AdS_{5}\times S^{5}$ and $AdS_{3}\times S^{3}\times T^{4}$, thus
we expect an identical integrable description. However at one loop
the string should feel the entire space: this is seen explicitly in
the list of modes needed for semiclassical analysis, and is encoded
in the dressing phase of integrability.

The circular spinning string in $S^{3}$ which we study is given by
\cite{Arutyunov:2003za} 
\begin{equation}
t=\kappa\tau,\qquad Z_{1}=\tfrac{1}{\sqrt{2}}e^{i(\mathcal{J}\tau+m\sigma)},\qquad Z_{2}=\tfrac{1}{\sqrt{2}}e^{i(\mathcal{J}\tau-m\sigma)}\label{eq:class-soln-ART}
\end{equation}
where $t$ is $AdS$ time, $S^{3}\subset\mathbb{C}^{2}$ has co-ordinates
$Z_{i}$, and we take $\sigma\in[0,2\pi]$. Clearly $m\in\mathbb{Z}$
is a winding number. The Virasoro constraints impose $\kappa=\sqrt{\cJ^{2}+m^{2}}$.
This solution has two equal angular momenta $J_{1}=J_{2}=R^{2}\cJ/2$,\footnote{Note the factor of 2. If we write $J=J_{1}+J_{2}$ then $\cJ=J/\sqrt{\lambda}$.
Later we use $J=L$, for instance in \eqref{eq:defn-gHL}.} and energy $\Delta=R^{2}\kappa$. We define the 't Hooft coupling
as in $AdS_{5}$ by $R^{2}=\sqrt{\lambda}$. 

This solution is particularly simple in the string theory because
it is homogeneous, in the sense that a translation along $\sigma$
maps to an isometry of the target space. It is also particularly simple
in integrability because all the Bethe roots lie on one connected
curve. Below we present the calculation of its energy at one loop
in both of these pictures.

\subsection{Worldsheet semiclassical calculation\label{sub:Worldsheet-semiclassical}}

The bosonic modes of the same solution in $S^{5}$ were calculated
in \cite{Frolov:2004bh}. While it is simple to repeat their calculation,
there is no need to do so, as we can safely just keep the modes lying
in $S^{3}$, and discard the other two. They have frequencies
\[
w_{n}^{S\pm}=\sqrt{n^{2}+2\cJ^{2}\pm2\sqrt{n^{2}(\cJ^{2}+m^{2})+\cJ^{4}}}.
\]
The bosonic modes in AdS directions of course have mass $s=\kappa$,
and those in torus directions are massless: 
\[
w_{n}^{A}=\sqrt{n^{2}+\kappa^{2}},\qquad w_{n}^{T}=\left|n\right|.
\]

For the fermionic modes, the calculation was performed in \cite{Beccaria:2012kb},
and we briefly sketch it here. The equations of motion are as usual
given by  $\rho_{-}D_{+}\Theta^{1}=0=\rho_{+}D_{-}\Theta^{2}$ where
$D_{\mu}\Theta^{I}=(\partial_{\mu}+t_{\mu})\Theta^{I}+F\rho_{\mu}\Theta^{\text{\scriptsize not}I}$
with $\mu=0,1$, 
\[
\rho_{\mu}=\partial_{\mu}X^{M}E_{M}^{A}\Gamma_{A},\qquad t_{\mu}=\tfrac{1}{4}\partial_{\mu}X^{M}\omega_{M}^{AB}\Gamma_{AB},\qquad F=\tfrac{1}{4}\not F_{(3)}=\tfrac{1}{4}\left(\Gamma^{012}+\Gamma^{345}\right)
\]
and $\partial_{\pm}=\partial_{0}\pm\partial_{1}$, $\rho_{\pm}=\rho_{0}\pm\rho_{1}$
etc. If we adopt the $\kappa$-gauge $\Theta^{1}=\Theta^{2}$ then
the equations of motion simplify to 
\[
0=\left(\rho_{\mp}\partial_{\pm}+\rho_{\mp}t_{\pm}+\rho_{\mp}F\,\rho_{\pm}\right)\Theta(\sigma,\tau).
\]
Taking the sum of these equations, and using a plane wave ansatz for
the modes $\Theta(\sigma,\tau)=e^{iw_{n}\tau+in\sigma}\Theta_{0}$
(where $\Theta_{0}$ is a constant Majorana--Weyl spinor), we get
\[
0=\left(\rho_{0}\partial_{0}-\rho_{1}\partial_{1}+\rho_{0}t_{0}-\rho_{1}t_{1}+\rho_{0}F\,\rho_{0}-\rho_{1}F\,\rho_{1}\right)e^{iw_{n}\tau+in\sigma}\Theta_{0}.
\]
Solving this equation for the mode frequencies, we find equally many
massive and massless fermions: 
\[
w_{n}^{F}=\begin{cases}
\left|n\right| & \mbox{4 massless}\\
\sqrt{n^{2}+\cJ^{2}} & \mbox{4 of mass }\cJ.
\end{cases}
\]

With these mode frequencies we can now calculate the semiclassical
one-loop energy correction 
\[
\delta E=\sum_{n}e(n),\qquad e(n)=\sum_{b}^{8+8}(-1)^{F}\frac{1}{2\kappa}w_{n}^{b}.
\]
Clearly the 4 massless fermionic and 4 massless bosonic modes cancel,
and we need only the 4+4 massive mode frequencies. Using the re-summation
procedure of \cite{Beisert:2005cw}, the non-analytic term $\delta E^{\text{int}}$
was determined in \cite{Beccaria:2012kb} to be the following: 
\begin{equation}
\delta E_{\blmt}=\frac{m^{4}}{2\cJ^{3}}-\frac{7m^{6}}{12\cJ^{5}}+\frac{29m^{8}}{48\cJ^{7}}-\frac{97m^{10}}{160\cJ^{9}}+\frac{2309m^{12}}{3840\cJ^{11}}+\ldots\,.\label{eq:dE-BLMT}
\end{equation}
It is this non-analytic term that should be directly compared to the
Bethe ansatz calculation, presented next.

\subsection{Bethe Ansatz calculation\label{sub:Bethe-Ansatz-calculation}}

The relevant Bethe equations were found in \cite{Borsato:2013qpa},
and the $su(2)$ sector is exactly the same as the $su(2)$ sector
Bethe equation from $AdS_{5}\times S^{5}$, as it must be. This is
the case $\eta=+1$ of 
\begin{equation}
\left(\frac{x_{i}^{+}}{x_{i}^{-}}\right)^{L}=\prod_{j\neq i}^{K}\;\left[\frac{x_{i}^{+}-x_{j}^{-}}{x_{i}^{-}-x_{j}^{+}}\right]^{\eta}\Bigg(\frac{1-\oneover{x_{i}^{+}x_{j}^{-}}}{1-\oneover{x_{i}^{-}x_{j}^{+}}}\Bigg)\sigma^{\twobb}(x_{i},x_{j})^{2}.\label{eq:BAE-eta}
\end{equation}
Here we use the following expansion of the dressing phase 
\begin{equation}
\sigma^{\twobb}(x,y)=\exp\Big[\frac{i}{4\pi}\sum_{r,s}c_{r,s}\:Q_{r}(x)Q_{s}(y)\Big],\qquad c_{r,s}=hc_{r,s}^{(0)}+c_{r,s}^{(1)}+\bigo{1/h}\label{eq:dressing-phase-QQ}
\end{equation}
with $\smash{c_{r,s}^{(0)}}$ the usual AFS phase \cite{Arutyunov:2004vx}.
We normalise the one-loop phase $c_{r,s}^{(1)}$ as in \cite{Hernandez:2006tk},
and note that in $AdS_{3}$ we are interested in the left-left phase.\footnote{The spectrum of $AdS_{3}\times S^{3}\times T^{4}$ divides massive
particles into left and right sectors. The product of the left-left
phase $\sigma^{\twobb}$ and the left-right phase $\tilde{\sigma}^{\twobb}$
is in fact the dressing phase of $AdS_{5}$. We assume that the coefficients
$c_{r,s}$ are antisymmetric in $r\leftrightarrow s$, and zero when
$r+s$ even. } 

The analysis we need is identical to that of \cite{Hernandez:2006tk},
and we briefly review the procedure. Assuming that there is just one
cut (with one mode number $k$), in the thermodynamic limit we can
replace the product by an integral. Multiplying by the relevant weight
and integrating over the cut, \cite{Hernandez:2006tk} obtained this
expression in terms of resolvents:\footnote{We write tildes on quantities defined to match \cite{Hernandez:2006tk},
which differ from those in the S-matrix and dressing phase papers
\cite{Borsato:2013qpa,Borsato:2013hoa,Abbott:2013mpa,Borsato:2014hja}
by powers of $\ghl$. The spectral parameters are related $\xhl\equiv\ghl\,x$.
The charges are related $\Qhl_{n}=\ghl^{n-1}Q_{n}$ but we always
use $\Qhl_{n}$ for the total, and $Q_{n}(x_{k}^{\pm})$ for the constituents. } 
\begin{align}
G^{2}-2\pi kG-\eta\Gsuper{1} & =\ghl^{2}\left[\Gsuper{1}{}^{2}-2\pi k\Gsuper{2}\right]+\ghl^{2}(1+\eta)\left[\Gsuper{1}\Qhl_{2}-\Gsuper{2}\Qhl_{1}\right]\nonumber \\
 & \qquad+\sum_{r,s}-2\,\eta\,c_{r,s}^{(1)}\frac{1}{\sqrt{\lambda}}\ghl^{r+s-1}\left[\Gsuper{r}\Qhl_{s}-\Gsuper{s}\Qhl_{r}\right]+\bigodiv{\lambda}\label{eq:BAE-integrated}
\end{align}
where
\[
G(\xhl)=-\sum_{n=0}^{\infty}\Qhl_{n+1}\xhl^{n},\qquad\Gsuper{r}=-\sum_{m=0}^{\infty}\Qhl_{m+r+1}\xhl^{m}
\]
and
\begin{equation}
\ghl=\frac{1}{4\pi\cJ}=\frac{h}{2L}.\label{eq:defn-gHL}
\end{equation}

The first line of \eqref{eq:BAE-integrated} is the classical Bethe
equation. To calculate the effect of the dressing phase (on the second
line), we perturb all the charges 
\[
\Qhl_{n}=\Qhl_{n}^{0}+\frac{1}{\sqrt{\lambda}}\:\delta\Qhl_{n},\qquad\delta\Qhl_{1}=0
\]
and expand to order $1/\sqrt{\lambda}$.\footnote{The Bethe equations give us an expansion in a coupling $h$ which
in general may be a nontrivial function of the 't Hooft coupling $\sqrt{\lambda}$.
However this does not happen in $AdS_{5}\times S^{5}$, nor (at least
to one loop) in $AdS_{3}\times S^{3}\times T^{4}$ where we have $\sqrt{\lambda}=2\pi h+\bigo{1/h}$.} We are interested in the simplest case with $k=2m$. This gives a
cancellation such that $\delta\Qhl_{2}$ can be found alone, and the
result is 
\[
\delta E=\frac{\ghl^{2}}{2\pi}\delta\Qhl_{2}=\sum_{r,s}c_{r,s}^{(1)}\ghl^{r+s}\frac{\Qhl_{r+1}^{0}\Qhl_{s}^{0}-\Qhl_{s+1}^{0}\Qhl_{r}^{0}}{\pi(1+2\ghl^{2}\Qhl_{2}^{0})}.
\]
Substituting in the charges from the one-cut resolvent 
\begin{equation}
G^{0}(\xhl)=2\pi m+\frac{\sqrt{1+(4\pi m\ghl)^{2}}-\sqrt{1+(4\pi m\xhl)^{2}}}{2(\xhl-\ghl^{2}/\xhl)}\label{eq:resolvent-G17}
\end{equation}
(starting with $\Qhl_{1}^{0}=-2\pi m$ and $\Qhl_{2}^{0}=\frac{1}{2\ghl^{2}}\left[-1+\sqrt{1+16\pi^{2}\ghl^{2}m^{2}}\right]$)
leads to 
\begin{equation}
\delta E=\frac{m^{4}c_{1,2}}{4\cJ^{3}}+\frac{m^{6}\left(-4c_{1,2}-c_{1,4}+c_{2,3}\right)}{16\cJ^{5}}+\frac{m^{8}\left(15c_{1,2}+5c_{1,4}+2c_{1,6}-5c_{2,3}-2c_{2,5}+c_{3,4}\right)}{64\cJ^{7}}+\ldots.\label{eq:dE-general-c_rs}
\end{equation}
Then using the coefficients from \cite{Hernandez:2006tk} (i.e. $c_{r,s}^{\text{HL}}=-8\frac{(r-1)(s-1)}{(r+s-2)(s-r)}$
for $r,s\geq2$) naturally gives us the $AdS_{5}$ answer:
\[
\delta E_{\text{HL}}=-\frac{m^{6}}{3\cJ^{5}}+\frac{m^{8}}{3\cJ^{7}}-\frac{49m^{10}}{120\cJ^{9}}+\frac{2m^{12}}{5\cJ^{11}}+\bigodiv{\cJ^{13}}.
\]
In fact the only change from the $AdS_{5}\times S^{5}$ derivation
made here is that we have allowed for the possibility of $c_{1,s}\neq0$
in \eqref{eq:dE-general-c_rs}. 

To use this result \eqref{eq:dE-general-c_rs} in $AdS_{3}\times S^{3}\times T^{4}$
we simply need to substitute in different coefficients. The first
set $c_{r,s}^{\blmt}$ proposed in \cite{Beccaria:2012kb} are 
\begin{equation}
c_{r,s}^{\blmt}=2\frac{s-r}{r+s-2},\qquad r+s\mbox{ odd},\;r,s\geq1\label{eq:coeff-BLMT}
\end{equation}
and using these, we recover $\delta E_{\blmt}$ of \eqref{eq:dE-BLMT}
above, by design: \cite{Beccaria:2012kb} performed exactly this comparison. 

The second set of coefficients $\smash{c_{r,s}^{\bosst}}$ were found
by solving symmetry conditions on S-matrix including crossing symmetry
\cite{Borsato:2013hoa}, as well as from a semiclassical calculation
involving giant magnon scattering \cite{Abbott:2013mpa}.\footnote{This calculation was also done by \cite{Beccaria:2012kb}, who omitted
a crucial twist in the algebraic curve. The correct integral was calculated
earlier by \cite{David:2010yg}, however they did not express it in
terms of the charges $Q_{n}$. The analogous calculations in $AdS_{5}\times S^{5}$
are \cite{Gromov:2007cd,Chen:2007vs}, in that case done after both
\cite{Hernandez:2006tk} and \cite{Janik:2006dc,Beisert:2006ib}. } These two techniques agree perfectly, and also agree with the near-BMN
scattering amplitude \cite{Sundin:2014sfa}. They give 
\begin{equation}
c_{r,s}^{\mathrm{\bosst}}=\left[2\frac{s-r}{r+s-2}-\delta_{r,1}+\delta_{1,s}\right],\qquad r+s\mbox{ odd},\;r,s\geq1\label{eq:coeff-BOSST}
\end{equation}
and using these in \eqref{eq:dE-general-c_rs} we obtain 
\begin{align}
\delta E_{\bosst} & =+\frac{m^{4}}{4\cJ^{3}}-\frac{13m^{6}}{48\cJ^{5}}+\frac{25m^{8}}{96\cJ^{7}}-\frac{311m^{10}}{1280\cJ^{9}}+\frac{1723m^{12}}{7680\cJ^{11}}+\ldots\nonumber \\
 & =\delta E_{\blmt}-\frac{m^{4}}{4\cJ^{3}}+\frac{5m^{6}}{16\cJ^{5}}-\frac{11m^{8}}{32\cJ^{7}}+\frac{93m^{10}}{256\cJ^{9}}-\frac{193m^{12}}{512\cJ^{11}}+\ldots\nonumber \\
 & =\delta E_{\blmt}+\frac{m^{2}(\cJ-\kappa)}{2\kappa^{2}}.\label{eq:dE-difference}
\end{align}
This difference is the mismatch we seek to explain. (The closed form
on the last line is guessed from the series, and checked to order
$1/\cJ^{15}$).

\section{The Contribution from Massless L\"{u}scher Terms\label{sec:The-Solution}}

When calculating quantum corrections to the mass of a particle, by
drawing Feynman diagrams for the self-energy, L\"{u}scher terms are
the effect of the new diagrams possible in finite volume, namely those
in which a loop wraps the space. The original context was relativistic
theories \cite{Luscher:1985dn,Klassen:1990ub}, for which (in $1+1$
dimensions) the effect is 
\begin{equation}
\delta E^{F}=-s\int\frac{d\theta}{2\pi}e^{-sL\cosh\theta}\cosh\theta\left[S_{ab}^{ab}(\theta+i\tfrac{\pi}{2})-1\right].\label{eq:luescher-relativistic}
\end{equation}
In this case the dispersion relation is $\varepsilon(p)=\smash{\sqrt{p^{2}+s^{2}}}=s\cosh\theta$
and the S-matrix is a function only of the difference of rapidities
$S(p_{a},p_{b})=S(\theta_{a}-\theta_{b})$. The reason the S-matrix
appears is that taking large $L$ puts the particle circling the space
on-shell, leaving just one integral over the loop momentum. This has
made the formula very useful for integrable theories, where the same
S-matrix is what defines the Bethe equations. The derivation of $\delta E^{F}$
can be done allowing arbitrary dispersion relations \cite{Janik:2007wt},
including the magnon dispersion relation of AdS/CFT integrability.
This has provided various tests at strong coupling \cite{Janik:2007wt,Gromov:2008ie,Arutyunov:2006gs,Astolfi:2007uz,Hatsuda:2008gd,Minahan:2008re}
including some in $AdS_{3}$ \cite{Abbott:2012dd}.

L\"{u}scher corrections are usually exponentially suppressed in large
volume $L$. The crucial observation for this paper is that the exponential
$\sim e^{-sL}$ in \eqref{eq:luescher-relativistic} contains the
mass $s$ of the virtual particle, and thus in a system with massless
particles, we no longer expect this suppression. There are similar
terms in which the particle wraps the space $n$ times, typically
$\sim e^{-n\,sL}$ and thus subleading. But with a massless virtual
particle, we expect these to all be of the same order.\footnote{For our non-relativistic system the exponent is not proportional to
the mass, but the conclusions of this paragraph still hold. See \eqref{eq:exponent-with-mod-q}
for the form: $\qstar\propto\left|q\right|$ means that the contribution
from near to $q=0$ in the integral is not exponentially suppressed. } 

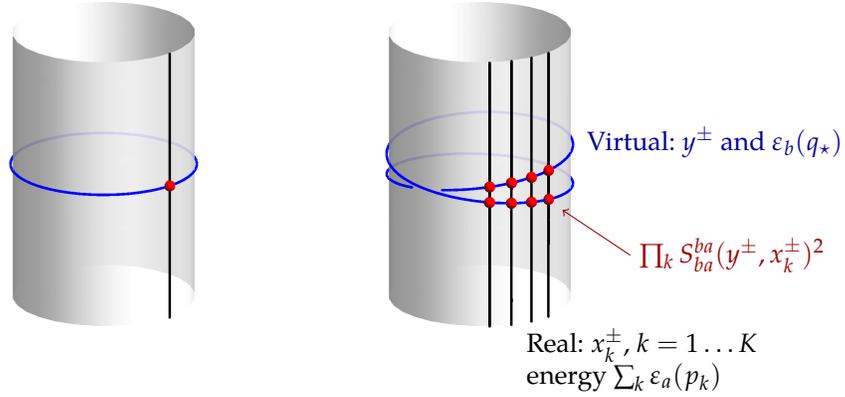
\begin{figure}
\vspace{-1cm}\centering 

\begin{tikzpicture}[scale=1]

\node at (-2,0) { \includegraphics[width=7cm]{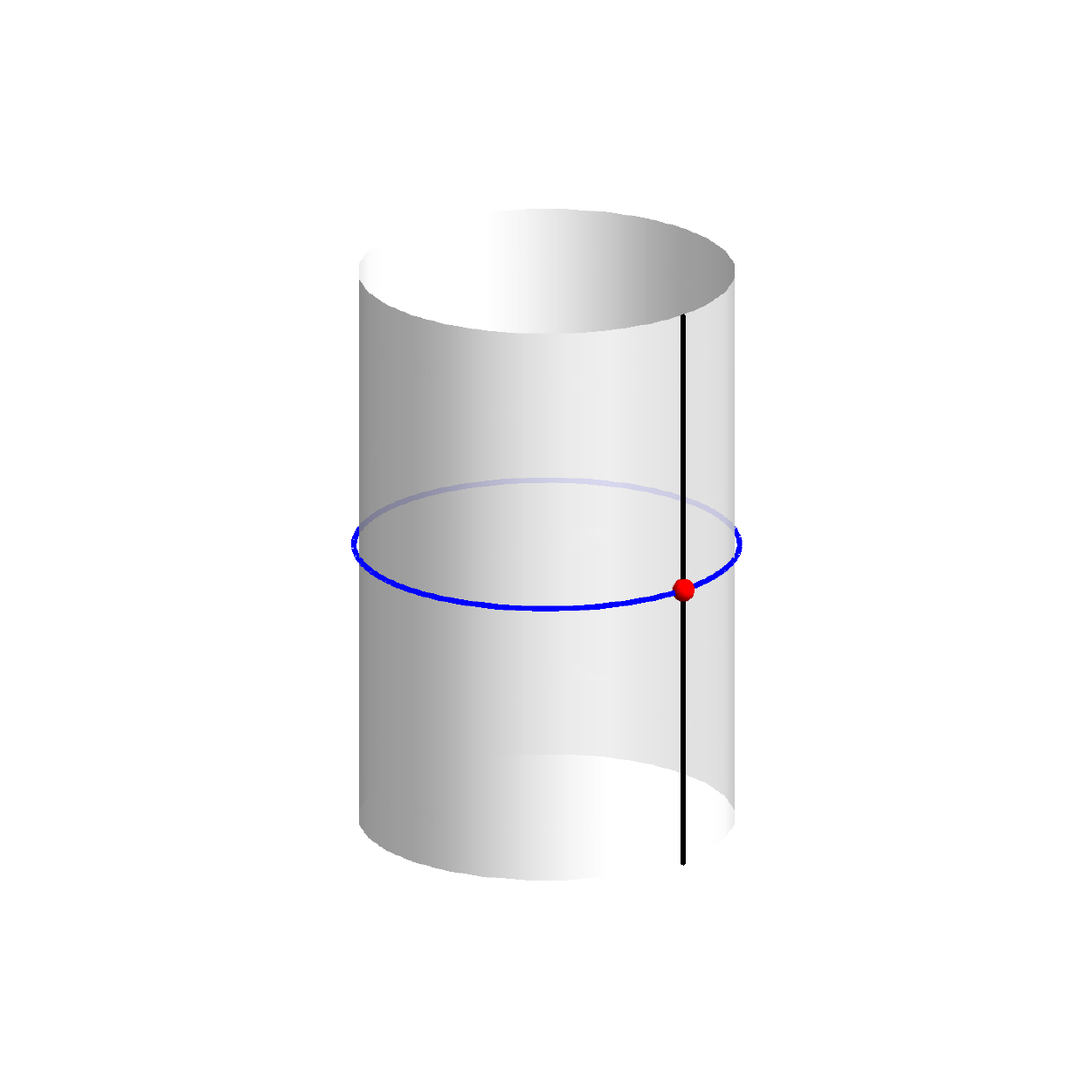} };
\node at (3,0) { \includegraphics[width=7cm]{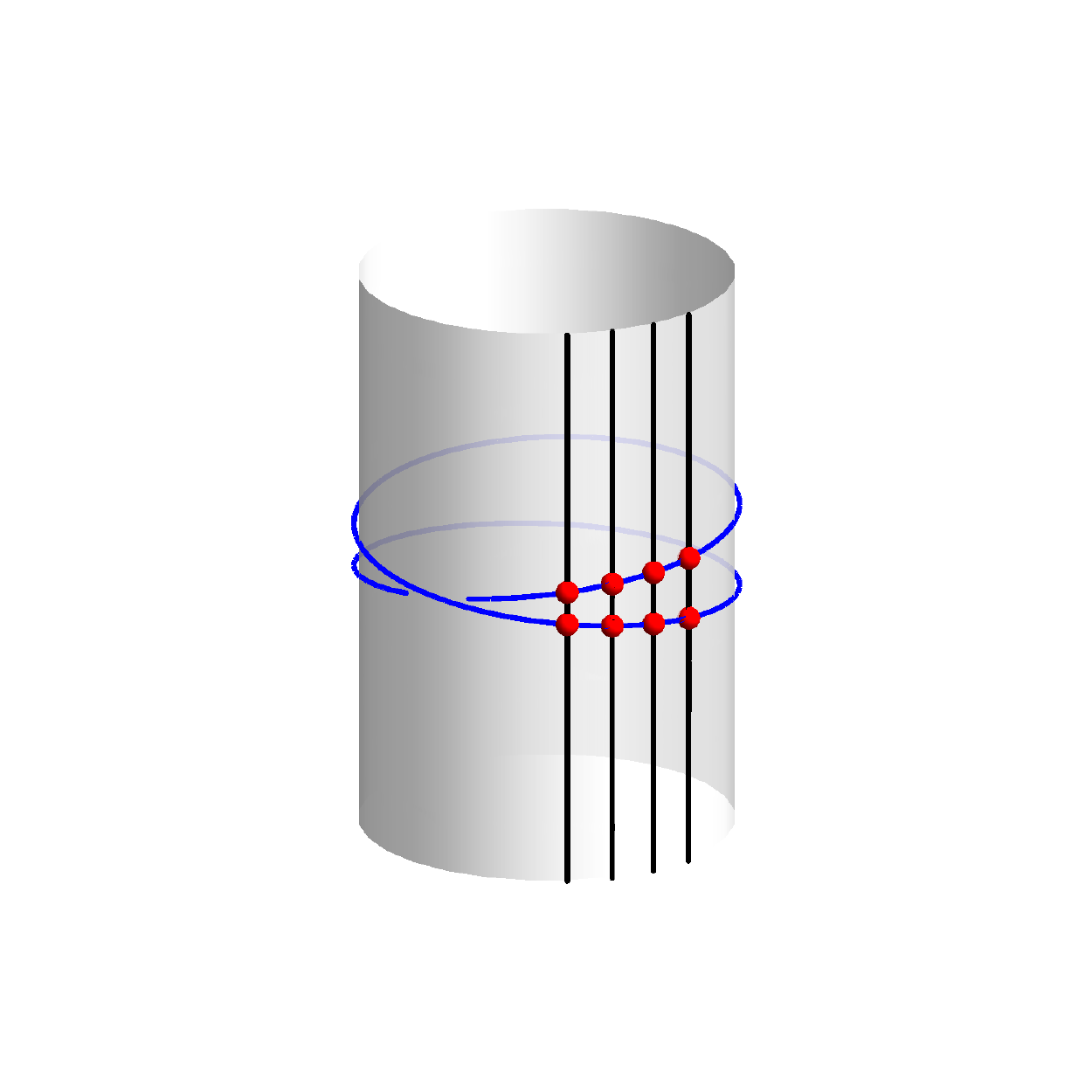} };

\node [darkblue, right, align=left] at (4.3,0.3) { Virtual: $y^\pm$ and  $\varepsilon_b(\qstar)$ };

\node [black, right, align=left] at (3.5,-2.6) { Real: $x_k^\pm$, $k=1\ldots K$ \\ 
  energy $ \sum_k \varepsilon_a(p_k)$ };

%\node [black, right, align=left] at (-1.5,-2.4) { Real: $x^\pm$ and $\varepsilon_a(p)$ };

\draw [darkred,<-] (4.1,-0.6) -- (5,-1.2) node [right, align=left] {$  \prod_k S_{ba}^{ba}(y^\pm,x_k^\pm)^2$};

%\draw [darkred,<-] (-1,-0.5) -- (0,-1.5) node [right, align=left] { $S_{ba}$};

\end{tikzpicture}

\vspace{-0.5cm}

\caption{On the left, the simplest F-term Feynman diagram. On the right, the
generalisations we consider allow a multiparticle physical state ($K=4$
shown) and multiple wrappings by the virtual particle ($n=2$ shown).
\label{fig:feynman+multimultiple}}
\end{figure}

Thus we need a generalisation of the simplest formula in two directions:
to treat a multiparticle physical state, and to allow multiple wrappings.
(Both are drawn in figure \ref{fig:feynman+multimultiple}.) These
have been studied separately in the literature, using techniques other
than the original Feynman diagrams, and were reviewed in \cite{Janik:2010kd}. 

\subsection{Multiply wrapped and multiparticle formulae}

The derivation of \cite{Heller:2008at} is a one-loop semiclassical
correction, treating the physical particle as a soliton. They extended
this to include the multiple wrappings appearing at one-loop level
(by picturing a cylinder of twice the radius with two physical solitons,
and so on), and obtained the following sum over $n$: 
\begin{equation}
\delta E_{\text{HJL}}=-\sum_{b}(-1)^{F_{b}}\fint_{-\infty}^{\infty}\frac{dq}{2\pi}\:\left(1-\frac{\varepsilon_{a}'(p)}{\varepsilon_{b}'(\qstar)}\right)\sum_{n=1}^{\infty}\frac{1}{n}\,e^{-in\qstar L}\left[S_{ba}^{ba}(\qstar,p)^{n}-1\vphantom{\frac{1}{1}}\right].\label{eq:dE-F-HJL}
\end{equation}
The $n=1$ term here agrees with \cite{Janik:2007wt}, and reduces
to \eqref{eq:luescher-relativistic} in the relativistic case. Pausing
to fix our notation, this is an energy correction to a particle of
type $a$ (and momentum $p$, dispersion relation $\varepsilon_{a}$)
due to virtual particles of all types $b$ (and $\varepsilon_{b}$)
circling the cylinder of size $L$. The momentum $q_{\star}$ is defined
as a function of $q$ by the on-shell condition $q^{2}+\varepsilon_{b}^{2}(\qstar)=0$.
The integration contour we use has the Euclidean energy $q$ real,
and thus $q_{\star}$ is imaginary. In this notation the Lorentzian
two-momenta of the real and virtual particles are 
\[
p_{\mu}=(\varepsilon_{a}(p),p)\qquad\mbox{and}\qquad q_{\mu}=(iq,\qstar)=\left(\varepsilon_{b}(\qstar),\qstar\right)\,.
\]

Another derivation is possible from the Thermodynamic Bethe Ansatz
(TBA); in fact this was the first approach in AdS/CFT \cite{Ambjorn:2005wa}.
A L\"{u}scher formula for multiparticle physical states was derived
in this way in \cite{Bajnok:2008bm}:
\begin{equation}
\begin{aligned}\delta E_{\text{BJ}} & =-\sum_{j,k}\varepsilon_{a}'(p_{k})\Big(\frac{\delta\mathrm{BY}_{k}}{\delta p_{j}}\Big)^{-1}\delta\Phi_{j}\\
 & \quad-\int_{-\infty}^{\infty}\frac{dq}{2\pi}\sum_{b_{1}\cdots b_{K}}(-1)^{F_{b_{1}}}\left[S_{b_{1}a}^{b_{2}a}(\qstar,p_{1})S_{b_{2}a}^{b_{3}a}(\qstar,p_{2})\cdots S_{b_{K}a}^{b_{1}a}(\qstar,p_{K})\right]e^{-i\qstar L}.
\end{aligned}
\label{eq:dE-F-multiparticle}
\end{equation}
 The pieces of first term come from writing the Bethe equations as
$2\pi n_{k}=\mathrm{BY}_{k}+\delta\Phi_{k}$ with 
\begin{align*}
\mathrm{BY}_{k} & =p_{k}L-i\sum_{j\neq k}\log S_{aa}^{aa}(p_{k},p_{j})\allowdisplaybreaks[0]\\
\delta\Phi_{j} & =\int_{-\infty}^{\infty}\frac{dq}{2\pi}\sum_{b_{1}\cdots b_{K}}(-1)^{F_{b_{1}}}\left[S_{b_{1}a}^{b_{2}a}(\qstar,p_{1})\cdots S_{b_{K}a}^{b_{1}a}(\qstar,p_{K})\right]e^{-i\qstar L}\frac{\partial}{\partial q}\log S_{b_{j}\:a}^{b_{j+1}a}(\qstar,p_{j}).
\end{align*}
When $K=1$ this reduces to the $n=1$ term of \eqref{eq:dE-F-HJL}.
For the case of interest here, we will see that the first line of
\eqref{eq:dE-F-multiparticle} will vanish, and the sum over various
internal choices in $S_{b_{1}a}^{b_{2}a}S_{b_{2}a}^{b_{3}a}\cdots S_{b_{K}a}^{b_{1}a}$
will turn out to be trivial, as the structure of the S-matrix forces
$b_{k}=b$ always. 

Combining features of \eqref{eq:dE-F-HJL} and \eqref{eq:dE-F-multiparticle},
we will also consider the following formula: 
\begin{equation}
\delta E=-\fint_{-\infty}^{\infty}\frac{dq}{2\pi}\sum_{n=1}^{\infty}\frac{1}{n}\,e^{-in\qstar L}\sum_{b}^{4+4}(-1)^{F_{b}}\left[\prod_{k=1}^{K}S_{ba}^{ba}(\qstar,p_{k})\right]^{n}.\label{eq:dE-F-ours}
\end{equation}
All terms $n$ here will contribute at the same order. As was pointed
out by \cite{Heller:2008at} about \eqref{eq:dE-F-HJL}, at $n>1$
this omits other multiply wrapped effects which would be expected
to contribute at the same order. 

After some preliminaries, we apply the singly wrapped formula \eqref{eq:dE-F-multiparticle}
to the circular string in section \ref{sub:Singly-wrapped-term},
and then look at multiple wrappings using \eqref{eq:dE-F-ours} in
section \ref{sub:Sum-over-all-wrapping}.

\subsection{Variables and S-matrix}

In order to use the known S-matrix \cite{Borsato:2014exa,Borsato:2014hja}
we must describe both particles with Zukhovski variables. Allowing
a generic mass $s$ these are defined by\footnote{The mass is normalised so that $s=1$ in $AdS_{5}\times S^{5}$, rather
than $\kappa$ as in section \ref{sub:Worldsheet-semiclassical}.
Note also that this equation fixes the definition of $h$ to match
\cite{Abbott:2013mpa,Borsato:2014hja,Abbott:2014pia} but not \cite{Borsato:2013qpa,Beccaria:2012kb,Borsato:2013hoa}.} 
\[
p_{z}=-i\log\frac{\zp}{\zm},\qquad s_{z}=\frac{h}{2i}\Big(\zp+\frac{1}{\zp}-\zm-\frac{1}{\zm}\Big)
\]
and describe dispersion relation 
\[
\varepsilon_{z}(p_{z})=\frac{h}{2i}\Big(\zp-\frac{1}{\zp}-\zm+\frac{1}{\zm}\Big)=\sqrt{s_{z}^{2}+4h^{2}\sin^{2}\frac{p}{2}}.
\]
It is often useful to define the spectral parameter $z$ by
\[
z^{\pm}+\frac{1}{z^{\pm}}=z+\frac{1}{z}\pm i\frac{s_{z}}{h}.
\]
For the real particles, we will write $x^{\pm}$ (or rather $x_{k}^{\pm}$)
in terms of $x$ defined like this, see \eqref{eq:xpm-expansion}
below. 

For massive virtual particles $y^{\pm}$, usually one transforms the
integral on $q$ into an integral on $y$ along the upper half unit
circle. But when this is massless, the variable $y$ is is not useful
as it approaches $1$ in the limit $s_{y}\to0$. Instead, we write
everything in terms of $q=-i\varepsilon(\qstar)$, finding 
\begin{equation}
y^{\pm}=1\pm\frac{\left|q\right|}{2h}+\frac{\left|q\right|^{2}}{8h^{2}}+\bigodiv{h^{3}}\label{eq:ypm-with-q}
\end{equation}
and
\begin{equation}
\qstar=-\frac{i}{h}\left|q\right|+\frac{i\left|q\right|^{3}}{24h^{3}}+\ldots\quad\Rightarrow\quad e^{-i\qstar L}=e^{-L\left|q\right|/h}+\bigodiv{h^{3}}.\label{eq:exponent-with-mod-q}
\end{equation}

The relevant matrix components of the S-matrix for massless-massive
scattering are given in appendix N of \cite{Borsato:2014hja}. Taking
the physical particle to be $a=Y_{p}^{L}$ (a left-sector sphere boson),
we are interested in the following matrix components: 
\begin{align}
\hat{S}_{\qstar,p} & =\hat{S}_{ba}^{ba}(y^{\pm},x^{\pm})=\hat{S}_{p,\qstar}^{\;-1}\nonumber \\
 & =(N_{p,\qstar}^{\twobw})^{-1}\begin{cases}
(A_{p,\qstar}^{LL}B_{p,\qstar}^{LL})^{-1} & \mbox{4 massless bosons, }b=T_{\qstar}^{\dot{a}a}\vspace{2mm}\\
(A_{p,\qstar}^{LL}A_{p,\qstar}^{LL})^{-1} & \mbox{2 massless fermions, }b=\chi_{\qstar}^{a}\\
(B_{p,\qstar}^{LL}B_{p,\qstar}^{LL})^{-1} & \mbox{2 massless fermions, }b=\tilde{\chi}_{\qstar}^{a}
\end{cases}\label{eq:S-matrix-ABN-three-rows}
\end{align}
where in (M.1) of \cite{Borsato:2014hja} we find 
\[
A_{p,\qstar}^{LL}\equiv1,\qquad B_{p,\qstar}^{LL}\equiv\sqrt{\frac{x^{-}}{x^{+}}}\frac{x^{+}-y^{+}}{x^{-}-y^{+}}.
\]
We will also need the normalisation factor from (O.2) of \cite{Borsato:2014hja}:
\[
N_{p,\qstar}^{\twobw}\equiv\sqrt{\frac{x^{-}}{x^{+}}}\sqrt{\frac{1-\oneover{x^{-}y^{+}}}{1-\oneover{x^{+}y^{-}}}}\sqrt{\frac{1-\oneover{x^{-}y^{-}}}{1-\oneover{x^{+}y^{+}}}}.
\]
We have defined this without the dressing phase, so in full we want
$S_{p,\qstar}=\hat{S}_{p,\qstar}\:(\sigma_{p,\qstar}^{\twobw})^{2}$.

Our calculation will be sensitive to the classical part of the mixed-mass
dressing phase $\sigma^{\twobw}$, which we take to be\footnote{The massive sector phase $\sigma^{\twobb}$ contains an AFS phase
of this form, with $W=2$. This is $c_{r,s}^{(0)}$ in \eqref{eq:dressing-phase-QQ}.} 
\begin{equation}
\sigma_{\text{AFS}}^{\twobw}(x^{\pm},y^{\pm})=\exp\Big\{ i\frac{h}{W}\sum_{r=2}^{\infty}\left[Q_{r}(x^{\pm})Q_{r+1}(y^{\pm})-Q_{r+1}(x^{\pm})Q_{r}(y^{\pm})\right]\Big\}\label{eq:AFS-sum-QQ}
\end{equation}
in terms of charges defined $Q_{1}(z)\equiv p_{z}=-i\log(\zp/\zm)$
and, for $n>1$, 
\[
Q_{n}(z)\equiv\frac{i}{n-1}\left[\frac{1}{(\zp)^{n-1}}-\frac{1}{(\zm)^{n-1}}\right].
\]
For now we leave the coefficient $W$ unfixed.\footnote{In \cite{Abbott:2014pia} we used the same form \eqref{eq:AFS-sum-QQ}
for the AFS phase for particles of mass $\alpha$ and $1-\alpha$
(in $AdS_{3}\times S^{3}\times S^{3}\times S^{1}$). The coefficient
there was $W_{xy}=4s_{x}s_{y}/(s_{x}+s_{y})$, which goes to zero
if $s_{x}=1$, $s_{y}=0$.}

\subsection{Ingredients for the $su(2)$ circular string\label{sub:Ingredients}}

The S-matrix part of \eqref{eq:dE-F-ours} involves a product over
all the physical particles $x_{k}^{\pm}$. We can re-write this as
an integral over the cut in the resolvent, in exactly the same way
as is done for the Bethe equations to derive \eqref{eq:BAE-integrated}
above: 
\begin{align}
S_{b}\equiv\prod_{k=1}^{K}S_{ba}(y^{\pm},x_{k}^{\pm}) & =\exp\left[\sum_{k=1}^{K}\log S_{ba}(y^{\pm},x_{k}^{\pm})\right]\nonumber \\
 & =\exp\left[L\int_{C}d\xhl\:\frac{\xhl^{2}-\ghl^{2}}{\xhl^{2}}\:\rho(\xhl)\:\log S_{ba}\!\left(y^{\pm},x^{\pm}(\xhl)\right)\right].\label{eq:how-to-integrate}
\end{align}
The density $\rho$ encodes the resolvent as 
\[
G(\xhl)=\int_{C}d\zhl\:\frac{\rho(\zhl)}{\xhl-\zhl}=-\sum_{n=0}^{\infty}\Qhl_{n+1}\xhl^{n}
\]
and the integration is over the single connected cut $C$ defining
our solution. It is easy to work out $\rho$ from \eqref{eq:resolvent-G17},
but to do the integral over $\xhl$ it is much better to use identities
\[
\Qhl_{n}=\int_{C}d\zhl\:\frac{\rho(\zhl)}{\zhl^{n}}.
\]
The spectral parameters $\xhl,\zhl$ appearing here are scaled by
$\ghl$ relative to the $x^{\pm},y^{\pm}$ used in the S-matrix. This
follows the convention of \cite{Hernandez:2006tk}, and is convenient
for taking the limit $L\to\infty$ at fixed $\ghl$. This expansion
gives $x_{k}^{\pm}$ as follows: 
\begin{equation}
x^{\pm}=\frac{1}{\ghl}\left[\xhl\pm\frac{i}{2L}\:\frac{\xhl^{2}}{\xhl^{2}-\ghl^{2}}+\bigodiv{L^{2}}\right].\label{eq:xpm-expansion}
\end{equation}
The variables $y^{\pm}$ are still given by \eqref{eq:ypm-with-q}
above. We have an expansion in $h$, but using $h=2\ghl L$ we regard
this as an expansion in $L$, i.e. $y^{\pm}=1\pm\left|q\right|/4\ghl L+\bigo{1/L^{2}}.$

With these expansions we can now write the leading contribution for
the S-matrix terms 
\[
-\log B_{p_{k},\qstar}^{LL}=+\log N_{p_{k},\qstar}^{\twobw}=-\frac{i}{2L}\frac{\xhl}{(\xhl-\ghl)^{2}}+\bigodiv{L^{2}}.
\]
Expanding in $\ghl\ll1$ and integrating as in \eqref{eq:how-to-integrate},
we get (using that $\Qhl_{n}=0$ for all odd $n\geq3$) 
\begin{align}
2i\theta\equiv\sum_{k=1}^{K}-\log B_{p_{k},\qstar}^{LL} & =-\frac{i}{2}\Big(\Qhl_{1}+2\sum_{n=1}^{\infty}g^{n}\Qhl_{n+1}\Big)\nonumber \\
 & =-i\pi m+iG(\ghl).\label{eq:theta-sans-expansion}
\end{align}
We must perform a similar sum for the AFS phase, since \eqref{eq:AFS-sum-QQ}
refers to the constituent particles. The charges are 
\[
\begin{aligned}Q_{n}(x_{k}^{\pm}) & =\frac{\ghl^{n-1}}{L\,\xhl^{n-2}}\,\frac{1}{\xhl^{2}-\ghl^{2}}+\bigodiv{L^{2}}\\
Q_{n}(\ypm) & =-\frac{i\left|q\right|}{h}+\ldots
\end{aligned}
\]
and thus the total phase is 
\[
\prod_{k=1}^{K}\sigma_{\text{AFS}}^{\twowb}(y^{\pm},x_{k}^{\pm})=\prod_{k=1}^{K}\exp\left[-\frac{\left|q\right|}{W}Q_{2}(x_{k}^{\pm})\right]=\exp\left[-\frac{\left|q\right|\ghl\Qhl_{2}}{W}\right].
\]
Let us combine this with the $e^{-i\qstar L}$ factor as follows:
\begin{equation}
e^{-i\qstar L}(\sigma_{\text{AFS}}^{\twowb})^{2}=\exp\left(-\left|q\right|\phi\right),\qquad\phi=\frac{1}{2\ghl}+\frac{\ghl\Qhl_{2}}{W}.\label{eq:what-is-phi}
\end{equation}
Then finally putting all of this into the S-matrix \eqref{eq:S-matrix-ABN-three-rows},
we have 
\[
e^{-i\qstar L}S_{b}=\begin{cases}
e^{-\left|q\right|\phi} & b=T\mbox{ massless bosons}\vspace{2mm}\\
e^{-\left|q\right|\phi}e^{-2i\theta} & b=\mbox{\ensuremath{\chi}}\mbox{ massless fermions }\\
e^{-\left|q\right|\phi}e^{+2i\theta} & b=\tilde{\chi}.
\end{cases}
\]

\subsection{Singly wrapped term\label{sub:Singly-wrapped-term}}

Using all the pieces we have calculated, it is now straightforward
to work out the singly wrapped L\"{u}scher F-term \eqref{eq:dE-F-multiparticle},
or equivalently, the $n=1$ term of \eqref{eq:dE-F-ours}. We set
$W=2$, and assume $m$ is even when expanding in $\ghl=1/4\pi\cJ\ll1$:
\begin{align}
\delta E & =-\int_{-\infty}^{\infty}\frac{dq}{2\pi}\;\sum_{b}^{4+4}(-1)^{F_{b}}e^{-i\qstar L}S_{b}=-\int_{-\infty}^{\infty}\frac{dq}{2\pi}\;8e^{-\left|q\right|\phi}\sin^{2}\theta=-\frac{8}{\pi}\frac{\sin^{2}\theta}{\phi}\nonumber \\
 & =\frac{-m^{4}}{2\cJ^{3}}+\frac{15m^{6}+\pi^{2}m^{8}}{24\cJ^{5}}-\frac{990m^{8}+135\pi^{2}m^{10}+2\pi^{4}m^{12}}{1440\cJ^{7}}+\bigodiv{\cJ^{9}}.\label{eq:dE-first}
\end{align}
This is clearly of the same order as \eqref{eq:dE-difference}, although
all the coefficients are wrong (and there are unwanted powers of $m$).
These will be changed by including the $n>1$ terms below. But first
we check our claim above that this $n=1$ term of \eqref{eq:dE-F-ours}
agrees exactly with the multiparticle formula \eqref{eq:dE-F-multiparticle}: 
\begin{itemize}
\item We can see that the first line of \eqref{eq:dE-F-multiparticle} vanishes
by calculating $\delta\Phi_{j}$. The constituent S-matrix elements
are independent of $q$, and thus $\smash{\frac{\partial}{\partial q}\log S_{b_{j}\:a}^{b_{j+1}a}(\qstar,p_{j})=0}$. 
\item We have simplified $S_{b_{1}a}^{b_{2}a}S_{b_{2}a}^{b_{3}a}\cdots S_{b_{K}a}^{b_{1}a}$
by setting $b_{k}=b$ for all $k$. Looking at the S-matrix as given
in (N.3) of \cite{Borsato:2014hja} again, notice that acting on $\ket{Y^{L}T^{\dot{a}a}}$
this never gives $\ket{X\:Y^{L}}$ with $X\neq T^{\dot{a}a}$. The
same is true for the massless fermions, thus no diagrams with $b_{k}\neq b_{k+1}$
survive.
\end{itemize}

\subsection{Sum over all wrappings\label{sub:Sum-over-all-wrapping}}

Doubly wrapped L\"{u}scher F-terms would usually (with massive particles)
be suppressed by the exponential factor squared. But this is not true
with massless virtual particles, and we find that the $n$-wrapped
terms all contribute at order $1/\cJ^{3}$, the same as the singly
wrapped term \eqref{eq:dE-first}. Thus we ought to sum all of them. 

We can do this using \eqref{eq:dE-F-ours}, which attempts to add
multiple wrapping corrections along the lines of \eqref{eq:dE-F-HJL}.
Using the same ingredients as above, the steps are as follows: \newcommand{\polylogtwo}{\mathop{\mathrm{Li}_2}\!}
\begin{align}
\delta E & =-\int_{-\infty}^{\infty}\frac{dq}{2\pi}\;\sum_{n=1}^{\infty}\;\sum_{b}^{4+4}(-1)^{F_{b}}\frac{1}{n}\left[e^{-i\qstar L}S_{b}\right]^{n}\nonumber \\
 & =-\int_{-\infty}^{\infty}\frac{dq}{2\pi}\;\sum_{n=1}^{\infty}\;\frac{8}{n}e^{-n\left|q\right|\phi}\sin^{2}n\theta\allowdisplaybreaks[0]\nonumber \\
 & =-\int_{-\infty}^{\infty}\frac{dq}{2\pi}\;2\log\frac{(1-e^{-\left|q\right|\phi+2i\theta})(1-e^{-\left|q\right|\phi-2i\theta})}{(1-e^{-\left|q\right|\phi})^{2}}\nonumber \\
 & =\frac{2}{\phi}\left[-\frac{\pi}{3}+\frac{1}{\pi}\polylogtwo\big(e^{2i\theta}\big)+\frac{1}{\pi}\polylogtwo\big(e^{-2i\theta}\big)\right].\label{eq:sum-in-n-luscher-half-way}
\end{align}
At this point we can expand in $\ghl$ by brute force,  but we can
also obtain the closed form of \eqref{eq:dE-difference} more elegantly.
Begin by observing that (for even $m$)\footnote{The case of odd $m$ should give the same physics, but one will need
to be more careful about branches of functions. } the term $\frac{-1}{4}Q_{1}=-\frac{1}{2}\pi m$ in $\theta$ \eqref{eq:theta-sans-expansion}
does not contribute to $\sin^{2}n\theta$. Thus we may replace $\theta$
with $\bar{\theta}$ given by
\[
\bar{\theta}=-\frac{2\pi^{2}m^{2}\ghl}{\sqrt{1+16\pi^{2}\ghl^{2}m^{2}}}=-\frac{\pi m^{2}}{2\kappa}
\]
recalling $\kappa=\sqrt{\cJ^{2}+m^{2}}$. We also write 
\[
\phi=\frac{2\pi}{W}\left[\cJ(W-1)+\kappa\right].
\]
Now we can use some properties of the dilogarithm to simplify \eqref{eq:sum-in-n-luscher-half-way}.
For $0<\left|\gamma\right|<2\pi$ we have \cite{Olver:2010-dilogchapter}
\[
\polylogtwo\big(e^{\pm i\gamma}\big)=\sum_{n=1}^{\infty}\frac{\cos\left(\pm n\gamma\right)}{n^{2}}+i\sum_{n=1}^{\infty}\frac{\sin\left(\pm n\gamma\right)}{n^{2}}.
\]
Then using the elementary sum $\sum_{n=1}^{\infty}\frac{\cos\left(\pm n\gamma\right)}{n^{2}}=\frac{\pi^{2}}{6}\mp\frac{\pi\gamma}{2}+\frac{\gamma^{2}}{4}$,
we obtain
\[
\polylogtwo\big(e^{+i\gamma}\big)+\polylogtwo\big(e^{-i\gamma}\big)=\frac{\pi^{2}}{3}+\frac{\gamma^{2}}{2}.
\]
For our calculation $\gamma=2\bar{\theta}$, and when $\ghl$ is small
($\kappa$ is large) this is in the required range. Thus \eqref{eq:sum-in-n-luscher-half-way}
becomes simply 
\begin{align}
\delta E & =\frac{1}{\pi\phi}\left(2\bar{\theta}\right)^{2}\nonumber \\
 & =-\frac{Wm^{2}\left[\cJ(W-1)-\kappa\right]}{2\kappa^{2}\left[1-(W-2)W\cJ^{2}/m^{2}\right]}\nonumber \\
 & =-\frac{m^{2}(\cJ-\kappa)}{\kappa^{2}}\qquad\mbox{ if }W=2.\label{eq:dE-result-exact}
\end{align}
We obtain the mismatch \eqref{eq:dE-difference} up to  a factor
of 2. Setting $W=2$ makes \eqref{eq:AFS-sum-QQ} precisely the usual
AFS phase.  Expanding, we get
\[
\delta E=-2\left[-\frac{m^{4}}{4\cJ^{3}}+\frac{5m^{6}}{16\cJ^{5}}-\frac{11m^{8}}{32\cJ^{7}}+\frac{93m^{8}}{256\cJ^{9}}+\ldots\right].
\]

The formula \eqref{eq:dE-F-ours} we used here is a rather crude attempt
to write down the appropriate multiply wrapped multiparticle L\"{u}scher
term. It is very encouraging that it almost works, but the real answer
is probably much more complicated. Some comments in this direction: 
\begin{itemize}
\item There is another derivation of L\"{u}scher terms from the TBA in
\cite{Ahn:2011xq,Bombardelli:2013yka}. The leading order (i.e. $n=1$)
term there $\delta E^{(1)}$ is \eqref{eq:dE-F-multiparticle}, but
they also derive two `next to leading order' (NLO) terms. One of these,
$\delta E^{(2,1)}$, is similar to our $n=2$ term but with a different
trace structure. The other, $\delta E^{(2,2)}$, has two momentum
integrals, and a factor of the virtual-virtual S-matrix. \medskip \\The
original application of these was at weak coupling. Naively attempting
to evaluate these terms at strong coupling, we do not see how to obtain
the correct results. For instance for giant magnons, we can compare
to \cite{Abbott:2011tp}, where we computed $n=2$ terms using \eqref{eq:dE-F-HJL}
and saw agreement with the string theory.
\item However it seems likely that a better derivation for terms with $n\geq2$
wrappings may include similar features: several loop momentum integrals,
and factors of the virtual-virtual S-matrix. While the $n=2$ term
may not be too difficult (from either Feynman diagrams or from TBA),
it seems clear that we will need to sum all $n$ for the effects studied
here. 
\end{itemize}
We stress that our conclusion that the Bethe ansatz is not complete
without massless L\"{u}scher corrections does not depend on these
details. The singly wrapped term \eqref{eq:dE-F-multiparticle} alone
produces a correction \eqref{eq:dE-first} of the same order as the
the mismatch \eqref{eq:dE-difference} which we set out to explain.

\section{Conclusion\label{sec:Conclusion}}

The problem we aimed to solve was that the correct one-loop dressing
phase $\sigma^{\twobb}$ (as deduced from crossing symmetry, and direct
semiclassical calculations) does not produce the correct one-loop
energy in the Bethe ansatz, for circular strings in $S^{3}$. The
solution we found is that L\"{u}scher F-terms with a massless virtual
particle circling the space contribute without an exponential suppression
factor, at precisely the right orders in $1/\cJ$ to repair this disagreement.
By summing over all wrappings, we are able to recover the difference
almost exactly. 

From this we conclude that the $su(2)$ Bethe equations are not sufficient
to describe the spectrum in this sector: they must be supplemented
by the effect of massless wrapping terms. This effect is the first
place in which the massless excitations of $AdS_{3}\times S^{3}\times T^{4}$
do not decouple from the massive sector. 

Our calculation involves the mixed-mass dressing phase $\sigma^{\twobw}$,
and we show that its classical term has the AFS form. We observe that
this phase is not the limit $\alpha\to1$ of the one needed for scattering
of mass $\alpha$ and mass $1-\alpha$ particles in the $AdS_{3}\times S^{3}\times S^{3}\times S^{1}$
case \cite{Abbott:2014pia}, and in particular cannot have the $Q_{1}Q_{2}$
term which seems to be necessary there.\footnote{The effect of using a classical phase starting with $Q_{1}Q_{2}$
as in \cite{Abbott:2014pia} in place of \eqref{eq:AFS-sum-QQ} is
that the final answer has a term $\delta E^{F}\sim1/\cJ^{4}$ which
is undesirable. In addition the coefficient $W_{xy}$ needed in \cite{Abbott:2014pia}
goes to zero as $s_{y}\to0$. } We interpret this as more evidence that the matching of the variables
used for that integrable system to those for $AdS_{3}\times S^{3}\times T^{4}$
is not simple. 

Massless L\"{u}scher terms will also be needed for macroscopic solutions
in $AdS_{3}\times S^{3}\times S^{3}\times S^{1}$, where the situation
is very similar: By placing the same resolvent in each $S^{3}$ one
finds an equally simple $su(2)$ sector of the Bethe equations, and
in the string theory one can likewise place the same solution in each
$S^{3}$ \cite{Beccaria:2012kb}. The correct massive dressing phase
there differs only by a factor of $\tfrac{1}{2}$ \cite{Abbott:2013mpa},
and the mixed-mass S-matrix is now also known \cite{Borsato:2015mma}.
Further ahead, similar effects are surely going to be important in
learning to treat `macroscopic massless' solutions in integrability,
as we proposed in \cite{Abbott:2014rca}.

\subsection*{Acknowledgements}

We thank Romuald Janik for many insightful comments and suggestions.
We have also benefitted from conversations with Diego Bombardelli,
Dmitri Sorokin, Per Sundin, Alessandro Torrielli, and Arkady Tseytlin
at various stages. 

Michael is supported by an NRF Innovation Fellowship. In\^{e}s was
supported by the NCN grant 2012/06/A/ST2/00396.

\bibliographystyle{my-JHEP-4}
\bibliography{complete-library-processed}

\end{document}